\documentclass[varenna]{cimento}
\usepackage{graphicx}  

\makeatletter
\def\alt{\mathrel{\mathpalette\vereq<}}
\def\vereq#1#2{\lower3pt\vbox{\baselineskip1.5pt \lineskip1.5pt
\ialign{$\m@th#1\hfill##\hfil$\crcr#2\crcr\sim\crcr}}}
\def\agt{\mathrel{\mathpalette\vereq>}}

\makeatother

\title{Astrophysical and Cosmological Neutrinos}
\author{Georg G.~Raffelt}

\begin{document}

\maketitle

\vskip-3.9cm

\noindent{\em Max-Planck-Institut f\"ur Physik 
(Werner-Heisenberg-Institut)\\
F\"ohringer Ring 6, 80805 M\"unchen, Germany}

\vskip 1cm

\begin{abstract}
We review the role of massive neutrinos in astrophysics and cosmology,
assuming that the oscillation interpretation of solar and atmospheric
neutrinos is correct. In particular, we discuss cosmological mass
limits, neutrino flavor oscillations in the early universe, 
leptogenesis, and neutrinos in core-collapse supernovae.
\end{abstract}

\vfill

\noindent Contribution to the Proceedings of the International School
of Physics ``Enrico Fermi,'' CLII Course ``Neutrino Physics,''
23 July--2 August 2002, Varenna, Lake Como, Italy. 
\eject


\section{Introduction}

The connection between neutrinos and astrophysics and cosmology is one
of the traditional pillars of astroparticle physics. On the one hand
side the intrinsic properties of neutrinos are difficult to measure;
the ``heavenly laboratories'' provide invaluable complementary
information~\cite{Raffelt1996,Raffelt1999,Dolgov2002a}.  On the other
hand side neutrinos dominate the dynamics of the radiation dominated
universe and of core-collapse supernovae and are important cooling
agents even for ordinary stars. Knowing the intrinsic neutrino
properties is crucial for our understanding of various astrophysical
and cosmological phenomena.

Yet the focus of neutrino astrophysics and cosmology is changing in
the light of what is beginning to be the established wisdom. Pure
laboratory experiments will soon overtake solar and atmospheric
neutrinos at measuring the mixing parameters.  While precision
cosmology continues to provide the most restrictive limit on neutrino
masses, the importance of astrophysics and cosmology as neutrino
laboratories is probably diminishing.  Rather, the measured neutrino
properties will be crucial input information in the astrophysical
context. For example, neutrinos are likely to become important as
``astrophysical messengers'' as the upcoming generation of
km$^2$-scale high-energy neutrino telescopes begins to open a new
window to the universe~\cite{Learned2000,Halzen2002}.  The measured
mixing parameters will be crucial for interpreting the neutrino
signals.

In my lectures I will focus on two other themes that connect neutrinos
with astrophysics.  One is the role of neutrino masses in cosmology.
The theory of structure formation in conjunction with recent and
upcoming galaxy redshift surveys provide the most restrictive limits
on neutrino masses (Sec.~\ref{sec:numass}).  Further, the anticipated
laboratory confirmation of the large-mixing-angle oscillation solution
of the solar neutrino problem in conjunction with big-bang
nucleosynthesis will for the first time fix the number density of
cosmic neutrinos and thus establish a unique connection between
neutrino masses and the cosmic hot dark matter fraction
(Sec.~\ref{sec:nunumber}).  Of course, massive neutrinos most likely
have less to do with cosmic dark matter than with the baryonic matter
by virtue of the leptogenesis mechanism for creating the baryon
asymmetry of the universe (Sec.~\ref{sec:leptogenesis}).

My second topic are supernova (SN) neutrinos.  The current and
upcoming large-scale neutrino detectors have a variety of primary
goals, but usually double as SN observatories.  The foreseeable
experiments may cover the neutrino sky for several decades so that one
may well expect an eventual SN observation. While the detection of
about 20 neutrinos from SN~1987A was a milestone of neutrino
astronomy, a high-statistics observation would allow one to extract a
wealth of information.  Section~\ref{sec:supernova} is devoted to the
mechanism of core-collapse SN explosions, the expected neutrino
signature, and the relevant detectors while Sec.~\ref{sec:SNlimits} is
given over to neutrino masses and oscillations.

The atmospheric neutrino anomaly indicates $\nu_\mu\to\nu_\tau$
oscillations \cite{Fukuda2000} with the mixing parameters shown in
table~\ref{tab:osci}.  The recent SNO results have largely established
active-active oscillations as a solution of the solar neutrino problem
\cite{Ahmad2002a,Ahmad2002b,Maltoni:2002ni}. The LMA parameters are
strongly favored, but the LOW case may still be viable.  There remains
the unconfirmed evidence for flavor transformations from the LSND
experiment~\cite{Eitel2000}. If interpreted in terms of oscillations,
the mixing parameters from all three sources are mutually
inconsistent. Even including a putative sterile neutrino no longer
provides a good global fit because all three cases prefer
active-active over active-sterile
oscillations~\cite{Strumia2002,Maltoni:2002xd}.  MiniBooNE at Fermilab
is expected to confirm or refute LSND within two
years~\cite{Tayloe2002}.  As there is no straightforward global
interpretation I will follow the widespread practice of discarding
LSND.  If it is due to neutrino conversions after all, something
fundamentally new is going on and much of what follows may have to be
revised.  I will always assume that there are three neutrino mass
eigenstates separated by the atmospheric and solar mass differences.
One obvious open question is the value of the overall neutrino mass
scale $m_\nu$. What is cosmology's contribution to clarifying this
issue?

\begin{table}
\caption{Experimental evidence for neutrino flavor
oscillations.\label{tab:osci}}
\begin{tabular}[4]{lllll}
\hline
\multicolumn{2}{l}{Evidence}
&Channel&$\Delta m^2$ [$\,\rm eV^2$]
&$\sin^22\Theta$\\
\hline
\multicolumn{2}{l}{Atmospheric}
&$\nu_\mu\to\nu_\tau$&(1.6--3.9)${}\times10^{-3}$&0.92--1\\
\multicolumn{1}{l}{Solar:}& \quad LMA
&$\nu_e\to\nu_{\mu\tau}$&
(0.2--2)${}\times10^{-4}$&0.2--0.6\\
&\quad LOW&$\nu_e\to\nu_{\mu\tau}$
&$1.3\times10^{-7}$
&0.92\\
\multicolumn{2}{l}{LSND}
&$\bar\nu_\mu\to\bar\nu_e$&0.2--10&(0.2--$3)\times10^{-2}$\\
\hline
\end{tabular}
\end{table}


\section{Neutrino Dark Matter and Cosmic Structure Formation}

\label{sec:numass}

The cosmic number density of neutrinos and anti-neutrinos per flavor
is $n_{\nu\bar\nu}=\frac{3}{11}\,n_\gamma$ with $n_\gamma$ the number
density of cosmic microwave photons, and assuming that a possible
asymmetry between the $\nu$ and $\bar\nu$ distributions is
negligible~\cite{Kolb1990}.  With $T_\gamma=2.728~{\rm K}$ this
translates into $n_{\nu\bar\nu}=112~{\rm cm}^{-3}$. If neutrinos have
masses one finds a cosmic mass~fraction
\begin{equation}\label{eq:Oega_nu}
\Omega_{\nu\bar\nu}h^2=\sum_{\rm flavors}\frac{m_\nu}{92.5~{\rm eV}}\,,
\end{equation}
where $h$ is the Hubble constant in units of $100~\rm
km~s^{-1}~Mpc^{-1}$.  The requirement that neutrinos not overclose the
universe implies the traditional limit $\sum m_\nu\alt 40~{\rm eV}$,
an argument that was first advanced in a classic paper by Gershtein
and Zeldovich~\cite{Gershtein1966}.

Later Cowsik and McClelland~\cite{Cowsik1973} speculated that massive
neutrinos could actually provide the dark matter in clusters of
galaxies.  However, it soon became clear that neutrinos were not a
good dark matter candidate for two reasons.  One is the limited phase
space for neutrinos gravitationally bound to a
galaxy~\cite{Tremaine1979}.  Therefore, galactic dark-matter neutrinos
must obey a {\em lower\/} mass limit of some 30~eV for typical
spirals, and even 100--200~eV for dwarf galaxies
(``Tremaine-Gunn-limit'').  Today the most restrictive laboratory
limits on the overall neutrino mass scale come from the
Mainz~\cite{Weinheimer1999} and Troitsk~\cite{Lobashev1999} tritium
experiments. The current limit is~\cite{Weinheimer2002}
\begin{equation}\label{eq:tritiumlimits}
m_\nu<2.2~{\rm eV}\qquad\hbox{at 95\% CL}\,.
\end{equation}
It applies to all mass eigenstates if the mass differences are given
by the atmospheric and solar oscillation interpretation.
Equation~(\ref{eq:tritiumlimits}) is so restrictive that neutrinos as
galactic dark matter are completely ruled out.

Secondly, cosmic structure formation provides powerful information on
the neutrino mass fraction $\Omega_{\nu\bar\nu}$.  The observed
structure in the distribution of galaxies is thought to arise from the
gravitational instability of primordial density fluctuations. The
small masses of neutrinos imply that they stay relativistic for a long
time after their decoupling (``hot dark matter''), allowing them to
stream freely, thereby erasing the primordial density fluctuations on
small scales \cite{Kolb1990,Doroshkevich1980}.  While this effect does
not preclude neutrino dark matter, it implies a top-down scenario for
structure formation where large structures form first, later
fragmenting into smaller ones. It was soon realized that the predicted
properties of the large-scale matter distribution did not agree with
observations~\cite{White1983}.

Today it is widely accepted that the universe has critical density and
that its mass-energy inventory sports several nontrivial
components. Besides some 5\% baryonic matter (most of it dark) there
are some 25\% cold dark matter in an unidentified physical form and
some 70\% of a negative-pressure component (``dark energy'').  And
because neutrinos do have mass, they contribute at least 0.1\%.  This
fraction is based on $m_3=50~{\rm meV}$, the smallest value consistent
with atmospheric neutrino oscillations.

\begin{figure}[t]
\hbox to\hsize{\hss
\includegraphics[width=0.55\textwidth]{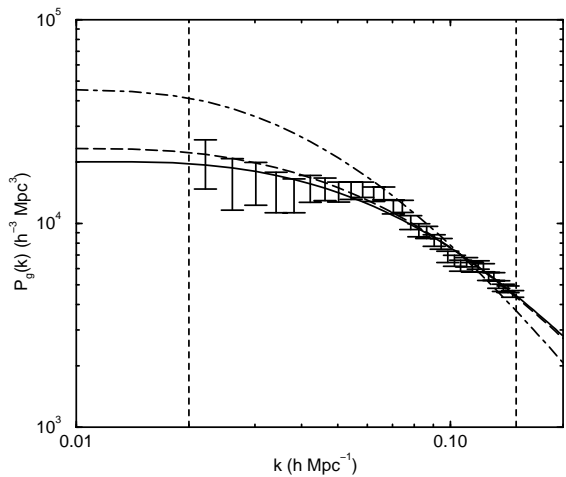}\hss}
\caption{Power spectrum of the galaxy distribution as measured by the
2dF Galaxy Redshift Survey. Theoretical predictions without neutrino
dark matter (solid line), for $\Omega_\nu=0.01$ (dashed), and
$\Omega_\nu=0.05$ (dot-dashed). The other cosmological parameters are
$\Omega_{\rm M}=0.3$, $\Omega_{\Lambda}=0.7$, $h=0.70$, and
$\Omega_{\rm B}h^2=0.02$.  (Figure from Ref.~\cite{Elgaroy2002} with
permission.)}
\label{fig:1}
\end{figure}

An upper limit on the neutrino dark matter fraction can be derived
from the power spectrum $P_{\rm M}(k)$ of the cosmic matter
distribution. Neutrino free streaming suppresses the small-scale
structure by an approximate amount \cite{Hu1998}
\begin{equation}
\frac{\Delta P_{\rm M}}{P_{\rm M}}\approx 
-8\,\frac{\Omega_\nu}{\Omega_{\rm M}}
\end{equation}
where $\Omega_{\rm M}$ is the cosmic matter fraction, excluding the
dark energy. This effect is illustrated in fig.~\ref{fig:1} where
$P_{\rm M}(k)$ measured by the 2dF Galaxy Redshift Survey is compared
with the predictions for a cold dark matter cosmology with neutrino
fractions $\Omega_\nu=0$, 0.01, and 0.05, respectively
\cite{Elgaroy2002}.
Based on the 2dFGRS data one finds $\sum m_\nu<1.8$--3.0~eV, depending
on the assumed priors for other cosmological parameters, notably the
Hubble constant, the overall matter fraction $\Omega_{\rm M}$, and the
tilt of the spectrum of primordial density fluctuations
\cite{Elgaroy2002,Hannestad2002}.  For a reasonable set of priors one
may adopt
\begin{equation}\label{eq:cosmicmasslimit}
\sum_{\rm flavors} m_\nu < 2.5~{\rm eV}
\end{equation}
at a statistical confidence level of 95\%. This limit corresponds
approximately to the dot-dashed ($\Omega_\nu=0.05$) curve in
fig.~\ref{fig:1}. Neutrinos may still contribute as much as 5\% of the
critical density, about as much as baryons.

Within the standard theory of structure formation, the largest
systematic uncertainty comes from the unknown biasing parameter $b$
that relates the power spectrum of the galaxy distribution to the
underlying matter distribution, $P_{\rm Gal}(k)=b^2\,P_{\rm
M}(k)$. The biasing parameter is one of the quantities which must be
taken into account when fitting all large-scale structure data to
observations of the galaxy distribution and of the temperature
fluctuations of the cosmic microwave background radiation. In the
future the Sloan Digital Sky Survey will have greater sensitivity to
the overall shape of $P_{\rm Gal}(k)$ on the relevant scales, allowing
one to disentangle more reliably the impact of $b$ and $\Omega_\nu$ on
$P_{\rm M}(k)$.  It is foreseen that one can then reach a sensitivity
of $\sum m_\nu \sim 0.65~{\rm eV}$ \cite{Hu1998}.

For degenerate neutrino masses eq.~(\ref{eq:cosmicmasslimit})
corresponds to a limit on the overall mass scale of $m_\nu<0.8~{\rm
eV}$, more restrictive than the laboratory limit
eq.~(\ref{eq:tritiumlimits}). However, the KATRIN project for
improving the tritium endpoint sensitivity is foreseen to reach
0.3~eV \cite{Osipowicz2001}, similar to the anticipated sensitivity of
future cosmological observations.  If both methods yield a positive
signature, they will mutually re-enforce each other. If they both find
upper limits, again they will be able to cross-check each other's
constraints.


\section{How Many Neutrinos in the Universe?}

\label{sec:nunumber}

The laboratory limits or future measurements of $m_\nu$ and of a hot
dark matter component can be related to each other if the cosmic
neutrino density $n_{\nu\bar\nu}$ is known. However, the cosmic
neutrinos can not be measured with foreseeable methods so that one
depends on indirect arguments for determining $n_{\nu\bar\nu}$.  Even
if we accept that there are exactly three neutrino flavors as
indicated by the $Z^0$ decay width and that they were once in thermal
equilibrium does not fix $n_{\nu\bar\nu}$.  Each flavor is
characterized by an unknown chemical potential $\mu_\nu$ or a
degeneracy parameter $\xi_\nu=\mu_\nu/T$, the latter being invariant
under cosmic expansion. While the small cosmic baryon-to-photon ratio
$\sim 10^{-9}$ suggests that the degeneracy parameters of all fermions
are very small, for neutrinos this is an assumption and not an
established fact.

In the presence of a degeneracy parameter $\xi_\nu$ the number and
energy densities of relativistic neutrinos plus anti-neutrinos of one
flavor are
\begin{eqnarray}
n_{\nu\bar\nu}&=&T_\nu^3\,\frac{3\zeta_3}{2\pi^2}\,
\left[1+\frac{2\ln(2)\,\xi_\nu^2}{3\zeta_3}
+\frac{\xi_\nu^4}{72\,\zeta_3}+O(\xi_\nu^6)\right]\,,\\
\rho_{\nu\bar\nu}&=&T_\nu^4\,\frac{7\pi^2}{120}\,
\left[1+\frac{30}{7}\left(\frac{\xi_\nu}{\pi}\right)^2
+\frac{15}{7}\left(\frac{\xi_\nu}{\pi}\right)^4\right]\,.
\end{eqnarray}
Therefore, if chemical potentials are taken to be the only
uncertainty, $n_{\nu\bar\nu}$ can only be larger than the standard
value. In this sense the structure-formation limits on the hot dark
matter fraction provide a conservative limit on $m_\nu$.  Conversely,
a laboratory limit on $m_\nu$ does not limit the hot dark matter
fraction.

Big-bang nucleosynthesis (BBN) is affected by $\rho_{\nu\bar\nu}$ in
that a larger neutrino density increases the primordial expansion
rate, thereby increasing the neutron-to-proton freeze-out ratio $n/p$
and thus the cosmic helium abundance. Therefore, the observed
primordial helium abundance provides a limit on $\rho_{\nu\bar\nu}$
which corresponds to some fraction of one effective extra neutrino
species. In addition, an electron neutrino chemical potential modifies
$n/p\propto\exp(-\xi_{\nu_e})$. Depending on the sign of $\xi_{\nu_e}$
this effect can increase or decrease the helium abundance and can
compensate for the effect of other flavors \cite{Kang1992}.  If
$\xi_{\nu_e}$ is the only chemical potential, BBN provides the limit
\begin{equation}
-0.01<\xi_{\nu_e}<0.07.
\end{equation}
Including the compensation effect, the only upper limit on the
radiation density comes from precision measurements of the power
spectrum of the temperature fluctuations of the cosmic microwave
background radiation and from large-scale structure measurements. A
recent analysis yields the allowed regions \cite{Hansen2002}
\begin{equation}
-0.01<\xi_{\nu_e}<0.22\,,\qquad |\xi_{\nu_{\mu,\tau}}|<2.6\,,
\end{equation}
in agreement with similar results of \cite{Hannestad2001} and
\cite{Kneller2001}.

However, the neutrino oscillations imply that the individual
flavor lepton numbers are not conserved and that in full thermal
equilibrium all neutrinos are characterized by one single chemical
potential $\xi_\nu$. If flavor equilibrium is achieved before $n/p$
freeze-out, the restrictive BBN limit on $\xi_{\nu_e}$ applies to all
flavors, i.e.\ $|\xi_\nu|<0.07$, implying that the cosmic number
density of neutrinos is fixed to within about 1\%. In that case the
relation between $\Omega_\nu$ and $m_\nu$ is uniquely given by the
standard expression eq.~(\ref{eq:Oega_nu}).

\begin{figure}[ht]
\hbox to\hsize{\hss
\includegraphics[width=0.55\textwidth]{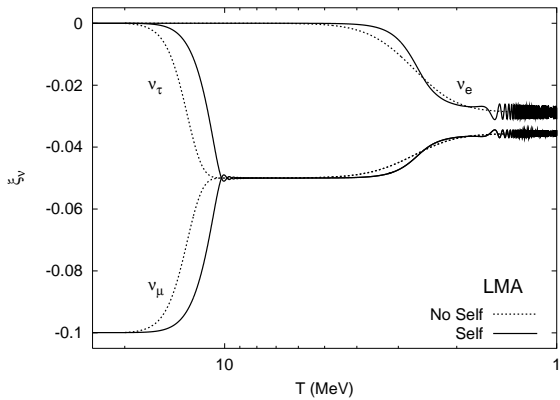}
\hss}
\caption{Cosmological evolution of neutrino degeneracy parameters for
the initial values $\xi_{\nu_e}=\xi_{\nu_\tau}=0$ and
$\xi_{\nu_\mu}=-0.1$.  The mixing parameters were chosen according to
the atmospheric and solar LMA solutions with $\Theta_{13}=0$.  (Figure
from Ref.~\cite{Dolgov2002b} with permission.)}
\label{fig:2}
\end{figure}

The approach to flavor equilibrium in the early universe by neutrino
oscillations and collisions was recently
studied~\cite{Lunardini2001a,Dolgov2002b,Wong2002,Abazajian2002}.
Assuming the atmospheric and solar LMA solutions, an example for the
cosmic flavor evolution is shown in fig.~\ref{fig:2}. The detailed
treatment is complicated and involves subtleties related to the large
weak potential caused by the neutrinos themselves. The intriguing
phenomenon of synchronized flavor
oscillations~\cite{Samuel1993,Pastor2002a} plays an interesting role,
that coincidentally could also be important in the context of
supernova neutrinos~\cite{Pastor2002b}.

The practical bottom line, however, is rather simple. Effective flavor
equilibrium before $n/p$ freeze-out is reliably achieved if the solar
oscillation parameters are in the favored LMA region. In the LOW
region, the result depends on the value of the small but unknown third
mixing angle $\Theta_{13}$. In the SMA and VAC regions, which are now
heavily disfavored, equilibrium is not achieved.  Establishing LMA as
the correct solution of the solar neutrino problem, for example by the
Kamland reactor experiment \cite{Shirai2002}, amounts in our context
to counting the number of cosmological neutrinos and thus to
establishing a unique relationship between $m_\nu$ and
$\Omega_{\nu\bar\nu}$.


\section{Leptogenesis}

\label{sec:leptogenesis}

Neutrino masses in the sub-eV range can play an interesting albeit
indirect role for creating the baryon asymmetry of the universe (BAU)
in the framework of leptogenesis scenarios \cite{Fukugita1986}.  The
main ingredients are those of the usual see-saw mechanism for small
neutrino masses. Restricting ourselves to a single family, the
relevant parameters are the heavy Majorana mass $M$ of the ordinary
neutrino's right-handed partner and a Yukawa coupling $g_\nu$ between
the neutrinos and the Higgs field $\Phi$.  The observed neutrino then
has a Majorana mass
\begin{equation}
m_\nu=\frac{g_\nu^2\langle\Phi\rangle^2}{M}
\end{equation}
that can be very small if $M$ is large, even if the Yukawa coupling
$g_\nu$ is comparable to that for other fermions.  Here,
$\langle\Phi\rangle$ is the vacuum expectation value of the Higgs
field which also gives masses to the other fermions.

The heavy Majorana neutrinos will be in thermal equilibrium in the
early universe. When the temperature falls below their mass, their
density is Boltzmann suppressed. However, if at that time they are no
longer in thermal equilibrium, their abundance will exceed the
equilibrium distribution. The subsequent out-of-equilibrium decays can
lead to the net generation of lepton number. CP-violating decays are
possible by the usual interference of tree-level with one-loop
diagrams with suitably adjusted phases of the various couplings. The
generated lepton number excess will be re-processed by standard-model
sphaleron effects which respect $B-L$ but violate $B+L$. It is
straightforward to generate the observed BAU by this mechanism.

The requirement that the heavy Majorana neutrinos freeze out before
they get Boltzmann suppressed implies an upper limit on the
combination of parameters $g_\nu^2/M$ that also appears in the see-saw
formula for $m_\nu$. The out-of-equilibrium condition thus implies an
upper limit on $m_\nu$.  Detailed scenarios for generic neutrino mass
and mixing schemes have been worked out, see
Ref.~\cite{Buchmuller2000} for a recent review and citations of the
large body of pertinent literature.  The bottom line is that neutrino
mass and mixing schemes suggested by the atmospheric and solar
oscillation data are nicely consistent with plausible leptogenesis
scenarios. Of course, it is an open question of how one would go about
to verify or falsify leptogenesis as the correct baryogenesis
mechanism. Still, it is intriguing that massive neutrinos may have a
lot more to do with the baryons than with the dark matter of the
universe!


\section{Core-Collapse Supernovae}

\label{sec:supernova}

\subsection{Explosion Mechanism}

\begin{figure}[ht]
\hbox to\hsize{\hss
\includegraphics[width=0.8\textwidth]{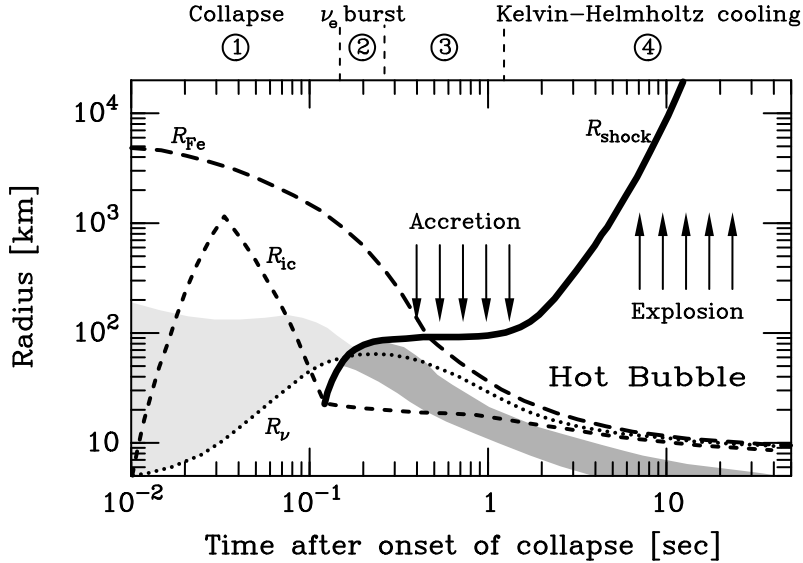}
\hss}
\caption{Schematic picture of the core collapse of a massive star
($M\protect\agt 8~M_{\odot}$) and the beginning of a SN
explosion. There are four main phases: 1.~Collapse. 2.~Prompt-shock
propagation and break-out, release of prompt $\nu_e$ burst. 3.~Matter
accretion and mantle cooling.  4.~Kelvin-Helmholtz cooling of
proto-neutron star.  The curves mark the evolution of several
characteristic radii: The stellar iron core ($R_{\rm Fe}$). The
neutrino sphere ($R_\nu$) with diffusive transport inside and free
streaming outside.  The ``inner core'' ($R_{\rm ic}$) for
$t\protect\alt0.1~{\rm s}$ is the region of subsonic collapse, later
it is the settled, compact inner region of the nascent neutron star.
The SN shock wave ($R_{\rm shock}$) is formed at core bounce,
stagnates for as much as several 100~ms, and then propagates outward.
The shaded area is the neutrino source region.}
\label{fig:3}
\end{figure}

A core-collapse supernova (SN) is perhaps the only system besides the
early universe that is dynamically dominated by neutrinos.  Moreover,
stellar collapse neutrinos have been observed once from SN~1987A and
may be observed again from a future galactic SN. A core-collapse SN
marks the evolutionary end of a massive star ($M\agt 8\,M_\odot$) that
has reached the usual onion structure with several burning shells, an
expanded envelope, and a degenerate iron core.  The core mass grows by
nuclear burning at its edge until it reaches the Chandrasekhar
limit. The collapse can not ignite nuclear fusion because iron is the
most tightly bound nucleus. Therefore, the collapse continues until
the equation of state stiffens at about nuclear density
($3\times10^{14}~{\rm g~cm^{-3}}$).  At this ``bounce'' a shock wave
forms, moving outward and expelling the stellar mantle and envelope,
i.e.\ the explosion is a reversed implosion.  Within the expanding
nebula, a compact object remains in the form of a neutron star or
perhaps sometimes a black hole. The kinetic energy of the explosion
carries about 1\% of the liberated gravitational binding energy of
about $3\times10^{53}~{\rm erg}$, the remaining 99\% going into
neutrinos.  The main phases of this sequence of events are illustrated
in fig.~\ref{fig:3}.

The bounce-and-shock explosion
mechanism~\cite{Brown1982,Woosley1986,Petschek1990,Burrows2000a} is
essentially a hydrodynamical phenomenon. However, realistic numerical
simulations have difficulties exploding for a physical reason.  The
shock wave forms {\it within\/} the iron core.  As it moves outward it
dissipates energy by the dissociation of iron.  The nuclear binding
energy of $0.1~M_\odot$ iron is about $1.7\times10^{51}~{\rm erg}$ and
thus comparable to the explosion energy. Therefore, the shock wave
stalls without driving off the stellar mantle and envelope.  The
standard scenario holds that the stagnating shock is ``re-juvenated''
by energy deposition so that enough pressure builds up behind the
shock to set it back into motion.  This ``delayed explosion scenario''
was first proposed in the early 1980s by Bethe and
Wilson~\cite{Bethe1984}.  One source of energy deposition behind the
shock wave is energy absorption from the nearly freely streaming
neutrinos which originate from the neutrino sphere near the
neutron-star surface.  Continued mass accretion and convection below
the shock wave also deposit energy and thus contribute to the shock
revival.

The main recent progress in numerical SN calculation has been the
implementation of efficient Boltzmann solvers so that a realistic
neutrino transport scheme can be self-consistently coupled with the
hydrodynamical evolution~\cite{Rampp2000,Mezzacappa2000,Burrows2000b}.
Such state-of-the-art spherically symmetric calculations do not lead
to successful explosions.  (The Livermore group does obtain robust
explosions~\cite{Totani1997}.  In their spherically symmetric
calculations they include a mixing-length treatment of ``neutron
finger convection,'' thereby enhancing the early neutrino luminosity
and thus the energy deposition behind the
shock~\cite{Wilson1993}. Their results agree with the findings of
other groups that diffusive neutrino transport alone is not enough to
trigger the explosion.) The spherically symmetric calculations are not
self-consistent in that the regions below the shock wave are
convectively unstable.  Forthcoming calculations will reveal if
convection, perhaps coupled with more accurate neutrino interaction
rates, will lead to successful explosions.

\subsection{Expected Neutrino Signal}

\begin{figure}[b]
\hbox to\hsize{\includegraphics[width=0.495\hsize]{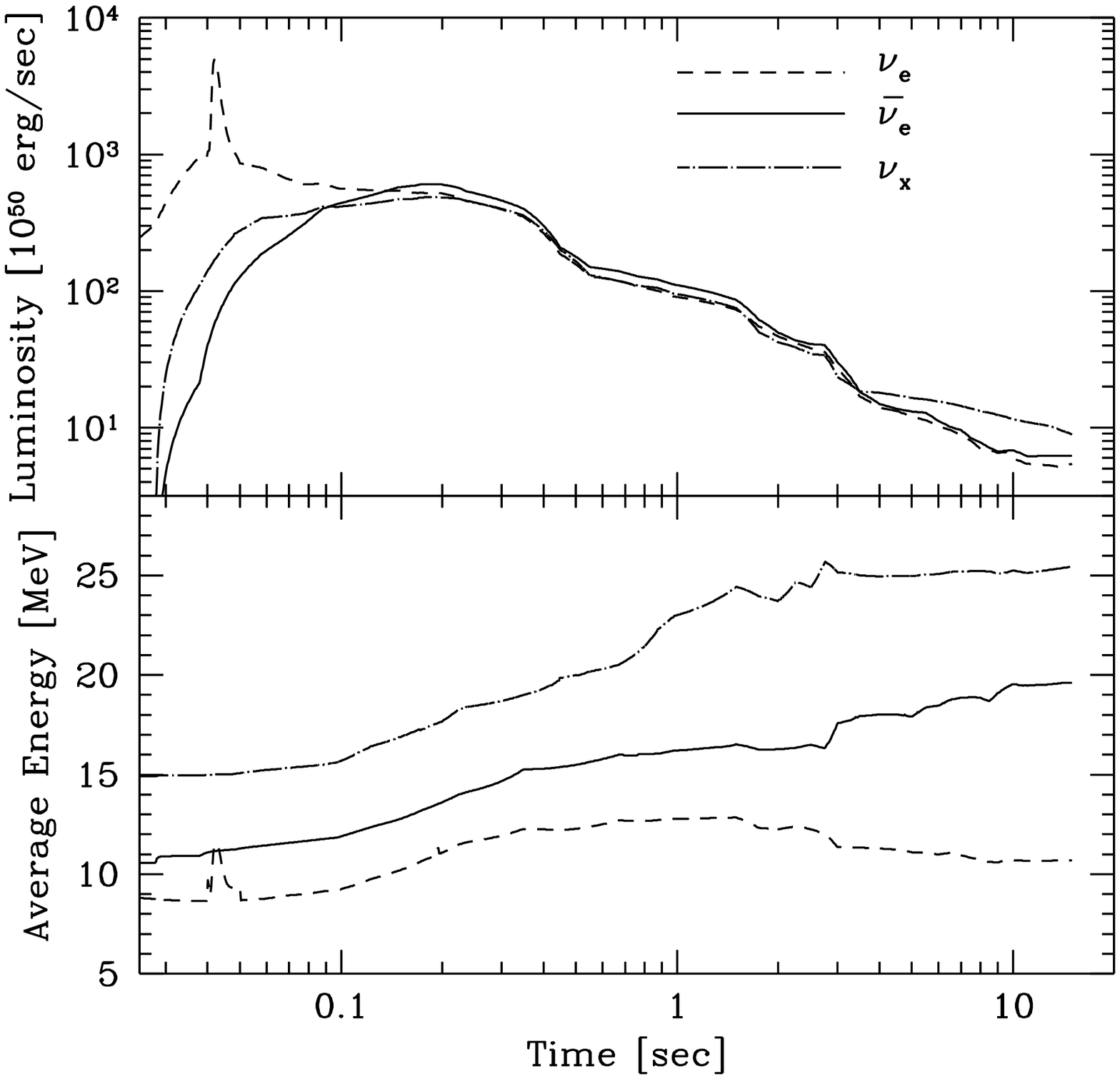}\hfil
\includegraphics[width=0.495\hsize]{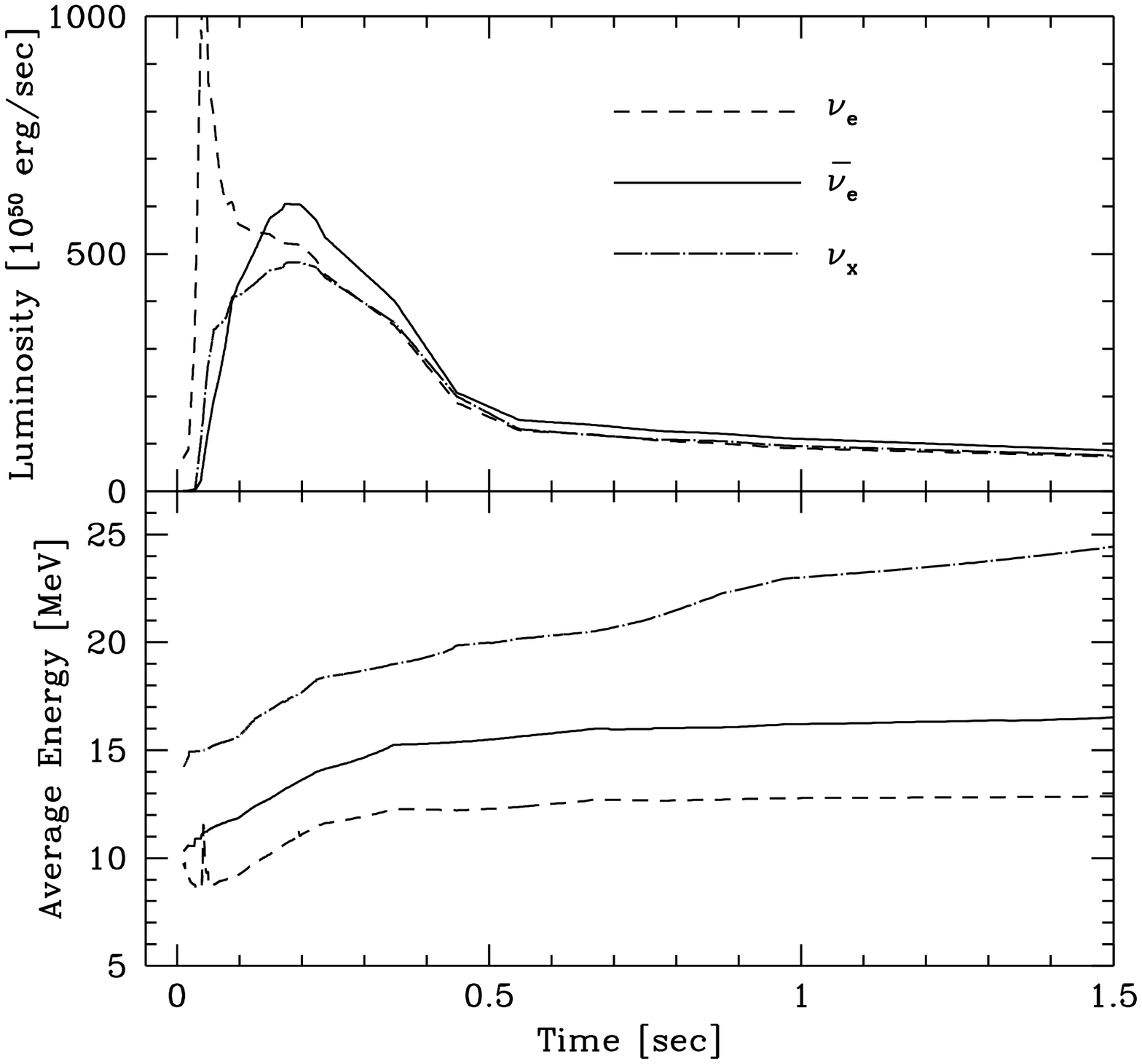}\kern-4pt}
\caption{SN neutrino luminosities and average energies from a
simulation with the Livermore code.  The $\nu_x$ line represents each
of $\nu_\mu$, $\bar\nu_\mu$, $\nu_\tau$ and $\bar\nu_\tau$.  Left
panel: Logarithmic luminosity and time scales.  Right panel: Linear
scales.  (Figures from Ref.~\cite{Totani1997} with permission.)}
\label{fig:4}
\end{figure}

The expected neutrino fluxes and spectra are illustrated by
fig.~\ref{fig:4}. The $\nu_e$ lightcurve shows a conspicuous spike
early on, representing the prompt neutrino burst which occurs when the
shock wave reaches the region of neutrino trapping in the iron
core. The dissociation of iron allows for the quick neutronization of
a layer of the proto neutron star. Of course, most of the lepton
number remains trapped and slowly escapes by neutrino diffusion.  The
subsequent broad shoulder up to about 300~ms, best visible in the
right panel with linear scales, represents the accretion phase where
material keeps falling in and powers the neutrino emission.  After
this phase the shock wave has driven off the stellar mantle.  The
subsequent long and flat tail represents the neutron star cooling by
neutrino emission.  The duration of the accretion phase depends on how
long it takes to revive the shock wave.  In the absence of a confirmed
robust explosion mechanism the exact duration of the accretion phase
is not known. Moreover, it may well depend on details of the star,
e.g.\ on the progenitor mass.

In order to observe effects of neutrino oscillations, the fluxes
and/or spectra must be different between the different flavors.  In
the simulation of fig.~\ref{fig:4} the neutrino luminosity is
virtually equipartitioned among the flavors after about 100~ms.  In
the recent Oakridge simulation \cite{Mezzacappa2000}, which includes a
state-of-the-art Boltzmann solver, the equipartition is also nearly
perfect between $\nu_e$ and $\bar\nu_e$, but the $\nu_x$ luminosity is
less than half that after 50~ms out to 600~ms when this simulation
terminates. Similar results are found in a Garching
simulation~\cite{Buras2002}.  Therefore, ``equipartition'' probably
should be taken to mean ``equal to within about a factor of two.''  A
recent systematic study of flavor-dependent neutrino transport also
concludes that one can not expect the flux equipartition to be exact
throughout the SN evolution~\cite{Keil2002}.

The neutrino average energies obey the well-known hierarchy $\langle
E_{\nu_e}\rangle <\langle E_{\bar\nu_e}\rangle <\langle
E_{\nu_x}\rangle$ which is explained by the different trapping
processes, $\beta$~processes for the electron flavor and elastic
scattering on nucleons for the rest. Therefore, the different flavors
originate in layers with different temperatures.  A physical
understanding of the neutrino spectra can be developed without
large-scale numerical simulations~\cite{Raffelt2001}.  While the
flavor hierarchy of average energies appears to be generic, the
differences are likely smaller than previously thought after all
relevant processes have been included, notably nucleon bremsstrahlung
and energy transfer by
recoils~\cite{Buras2002,Keil2002,Raffelt2001,Janka1995}.  The average
neutrino energies increase for the first few seconds. This is a
generic effect because the neutrino-emitting regions heat up by
accretion and by the contraction of the neutron star.

Numerical neutrino light curves can be compared with the SN~1987A data
where the measured energies are found to be ``too low.''  For example,
the numerical simulation of fig.~\ref{fig:4} yields time-integrated
values $\langle E_{\nu_e}\rangle\approx13~{\rm MeV}$, $\langle
E_{\bar\nu_e}\rangle\approx16~{\rm MeV}$, and $\langle
E_{\nu_x}\rangle\approx23~{\rm MeV}$.  On the other hand, the data
imply $\langle E_{\bar\nu_e}\rangle=7.5~{\rm MeV}$ at Kamiokande and
11.1~MeV at IMB~\cite{Jegerlehner1996}.  Even the 95\% confidence
range for Kamiokande implies $\langle E_{\bar\nu_e}\rangle<12~{\rm
MeV}$.  Flavor oscillations would increase the expected energies and
thus enhance the discrepancy~\cite{Jegerlehner1996}.  It has
remained unclear if these and other anomalies of the SN~1987A neutrino
signal should be blamed on small-number statistics, or point to a
serious problem with the SN models or the detectors.

\subsection{Possibility of a Future Detection}

\begin{figure}[t]
\hbox to\hsize{\hss
\includegraphics[width=0.55\hsize]{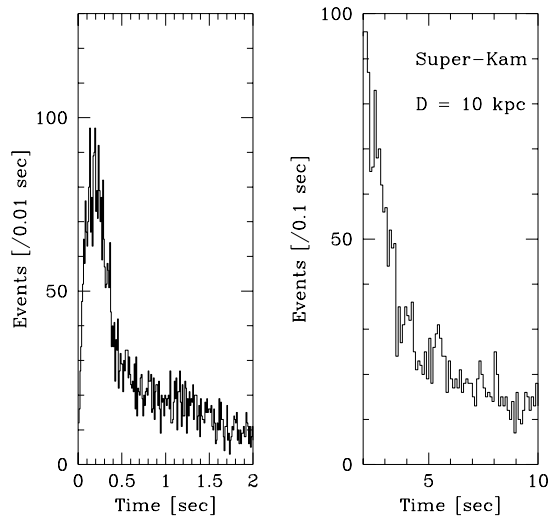}
\hss}
\caption{Simulated Super-Kamiokande neutrino light curve for
a galactic SN at 10~kpc, assuming the neutrino signal of
fig.~\ref{fig:4}. (Figure from Ref.~\cite{Totani1997}
with permission.)}
\label{fig:5}
\end{figure}

A serious confrontation of SN theory with neutrino data has to await a
high-statistics observation.  Detectors for measuring the neutrinos
from a galactic SN have almost continuously operated since 1980 when
the Baksan Scintillator Telescope (BST) started.  For a SN at a
distance of 10~kpc with neutrino fluxes and spectra roughly like those
of fig.~\ref{fig:4}, BST would register about 70~events.  The
neutrinos from SN~1987A in the Large Magellanic Cloud at a distance of
50~kpc were measured in Kamiokande~\cite{Hirata1988},
IMB~\cite{Bionta1987}, and BST~\cite{Alekseev1988} with a few events
each.  Today, much larger detectors exist, although BST keeps running.
Super-Kamiokande would measure about 8000 events from a SN at
10~kpc. A simulated light curve based on the SN model of
fig.~\ref{fig:4} is shown in fig.~\ref{fig:5}.  Of course, a SN
observation by Super-Kamiokande depends on the reconstruction of the
detector after its damage by the accident of 12
November~2001. Fortunately, there are other large observatories.  The
Sudbury Neutrino Observatory (SNO) would register about
800~events~\cite{Virtue2001}, ignoring flavor oscillations. The Large
Volume Detector (LVD) in the Gran Sasso Laboratory is a scintillation
detector that would register about 400~events~\cite{Aglietta2001}. A
similar number of events would be expected in the KamLAND
scintillation experiment which took up operations~\cite{Piepke2001}
and about 200 in MiniBooNE~\cite{Sharp2002}.  The Borexino solar
neutrino experiment, that will soon be ready, is smaller and would
register about 100~events~\cite{Cadonati2000}. The AMANDA South Pole
neutrino telescope also works as a SN neutrino detector in that the
correlated noise of all photomultipliers caused by the Cherenkov light
of the SN neutrinos produces a significant signal, especially when
AMANDA is enlarged to the km$^3$ IceCube~\cite{Ahrens2001}.

The dominant signal is usually the charged-current reaction $\bar\nu_e
p\to n e^+$.  SNO has a unique $\nu_e$ detection capability from the
CC deuterium dissociation $\nu_ed\to ppe^-$. Neutral-current reactions
which are sensitive to all flavors include elastic scattering on
electrons, the deuterium dissociation $\nu d\to np\nu$ in SNO, the
excitation of $^{16}\rm O$ in water Cherenkov detectors, and the
corresponding excitation of $^{12}\rm C$ in scintillation detectors,
notably in LVD and KamLAND, where the $\gamma$-rays from the
subsequent de-excitation can be measured.  Another possibility is
proton elastic scattering that can cause a measurable scintillation
signal~\cite{Beacom2002}.  Specific neutral-current detectors for SN
neutrinos have been proposed on the basis of the reaction
$\nu+(A,Z)\to (A-1,Z)+n+\nu$ where the neutron will be measured.  For
example, lead or iron could be used as targets in the proposed OMNIS
detector~\cite{Smith1997,Boyd2001}.  This sort of detector would be
complementary to Super-Kamiokande and SNO in that it is primarily
sensitive to the heavy-flavor neutrinos.

At present one debates the possibility of building even larger
detectors for precision neutrino long-baseline oscillation
experiments, proton decay, and high-statistics solar, atmospheric and
SN neutrino detection. A typical size could be a megatonne of water or
scintillator.  This option is discussed under the name of
Hyper-Kamiokande in Japan~\cite{Shiozawa2001}, under UNO in the
US~\cite{Vagins2001}, and is also debated in Europe~\cite{Frejus2002}.
Such a detector could produce as many as $10^5$ events from our
fiducial SN at 10~kpc.

The most crucial question is the SN rate in our galaxy because even
the largest foreseen detectors will barely cover the local group.  A
SN in the largest member, Andromeda (M31), at a distance of about
750~kpc would yield only about 30 events in a megatonne.  One approach
to estimating the galactic SN rate is to use the average rates derived
from external galaxies.  The most recent estimate is $2\pm1$
core-collapse SNe per century~\cite{Cappellaro2000}, about a factor
of~2 smaller than previous
estimates~\cite{vandenBergh1991,Tammann1994}.  Another approach relies
on the historical SN record, extrapolated to the entire
galaxy. Because of obscuration by dust, only SNe out to a few kpc have
been observed. The rate of core-collapse SNe is then estimted to be
3--4 per century~\cite{Tammann1994,Strom1994}, with a large Poisson
uncertainty from the small number of observed cases (5~SNe during the
past millenium).  Given the vagaries of small-number statistics, these
estimates agree with each other, and with circumstantial evidence such
as the estimated population of progenitor stars or the neutron-star
formation rate.

Large neutrino detectors are motivated by many goals so that it is not
unrealistic to expect another few decades of coverage.  Within that
time one may well observe a galactic SN even though they are rare. A
high-statistics neutrino light curve would allow one to observe
directly the collapse dynamics. For example, the early
accretion-powered neutrino emission could be clearly distinguished
from the subsequent neutron-star cooling phase~\cite{Totani1997}.  One
of the most energetic astrophysical phenomena would be caught in the
act, allowing one to unravel the underlying physics.


\section{Neutrino Masses and Oscillations}

\label{sec:SNlimits}

\subsection{Time-of-Flight Dispersion of Neutrino Burst}

In principle, neutrino masses can be measured by the dispersion of a
neutrino burst from a pulsed source, notably a SN. The time-of-flight
delay of massive neutrinos with energy $E_\nu$ is
\begin{equation}
\Delta t = \frac{m_\nu^2}{2E_\nu^2}\,D
\end{equation}
where $D$ is the distance to the source. Therefore, if a neutrino
burst has the intrinsic duration $\Delta t$ and the energies are
broadly distributed around some typical energy $E_\nu$, one is
approximately sensitive to masses
\begin{equation}
m_\nu > 10~{\rm eV}\,\left(\frac{E_\nu}{10~\rm MeV}\right)\,
\left(\frac{\Delta t}{\rm s}\right)^{1/2}
\left(\frac{10~\rm kpc}{D}\right)^{1/2}\,.
\end{equation}
The measured $\bar\nu_e$ burst of SN~1987A was characterized by
$E\approx 20~{\rm MeV}$, $\Delta t\approx 10~{\rm s}$, and $D\approx
50~{\rm kpc}$, leading to the well-known limit $m_{\nu}\alt20~{\rm
eV}$ \cite{Loredo1989}.  A recent re-analysis yields a somewhat more
restrictive limit~\cite{Loredo2002}.  Either way, these results are
only of historical interest because the tritium and cosmological
limits are now much more restrictive.

A high-statistics observation of a future galactic SN would yield more
restrictive limits because the relevant time scale $\Delta t$ is the
fast rise time of around 100~ms rather than the overall burst duration
of several seconds.  Therefore, one is sensitive to smaller masses
than the SN~1987A burst, despite the shorter baseline. Detailed
Monte-Carlo simulations imply that Super-Kamiokande would be sensitive
to about $m_{\nu}\agt3~{\rm eV}$, almost independently of the exact
distance~\cite{Totani1998}.  (At a larger distance one gains baseline
but loses statistics, two effects that virtually cancel for a given
detector size.)

Conceivably this sensitivity could be improved if a gravitational wave
signal could be detected preceding the neutrinos, signifying the
instant of the stellar collapse and bounce
\cite{Fargion1981,Arnaud2002}. In this case one may be sensitive to
about 1~eV.  It is also conceivable that a SN collapses to a black
hole some short time after the original collapse. In this case the
neutrino signal would terminate within $\Delta t\alt0.5~{\rm ms}$,
thereby defining a very short time scale.  Super-Kamiokande would be
sensitive to $m_{\nu}\agt1.8~{\rm eV}$ \cite{Beacom2000,Beacom2001}.
With a megatonne detector one could measure SN neutrinos throughout
the local group of galaxies. From Andromeda at a distance of 750~kpc
one would get around 30 events.  Using the overall signal duration for
$\Delta t$ yields a sensitivity of a few~eV.

The only conceivable time-of-flight technique that could probe the
sub-eV range involves Gamma-Ray Bursts (GRBs) which have been
speculated to be strong neutrino sources. If the neutrino emission
shows time structure on the millisecond scale, and assuming a
cosmological distance of 1~Gpc, one would be sensitive to neutrino
masses $m_\nu\agt 0.1~{\rm eV}~E_\nu/{\rm GeV}$. Therefore, observing
millisecond time structure in sub-GeV neutrinos from a GRB would be
sensitive to the sub-eV mass scale
\cite{Halzen1996,Choubey2002}.

\subsection{Flavor Oscillations}

Neutrino oscillations are now firmly established so that the SN
neutrino fluxes and spectra in a detector can be very different from
those emitted by the source. This is especially true if the favored
solar LMA case applies.  The relevant mass difference implies that
matter effects are important in the SN, and also in the Earth if the
neutrinos enter the detector ``from below.''  The large ``solar''
mixing angle $\theta_{12}$ implies that oscillations will be important
in both the $\nu_e$ and $\bar\nu_e$ channel.  In the LMA case it is
unavoidable, for example, that oscillations influence the SN~1987A
signal interpretation, and that the detectors saw different spectra
due to different Earth-crossing paths
\cite{Jegerlehner1996,Smirnov1994,Lunardini2000,Kachelriess2001}.
Taking oscillations into account will be crucial for extracting
information about the source if a future galactic SN is observed
\cite{Barger2001,Minakata2001,Schirato2002}.

\begin{figure}[ht]
\hbox{\hss
\includegraphics[width=0.5\hsize]{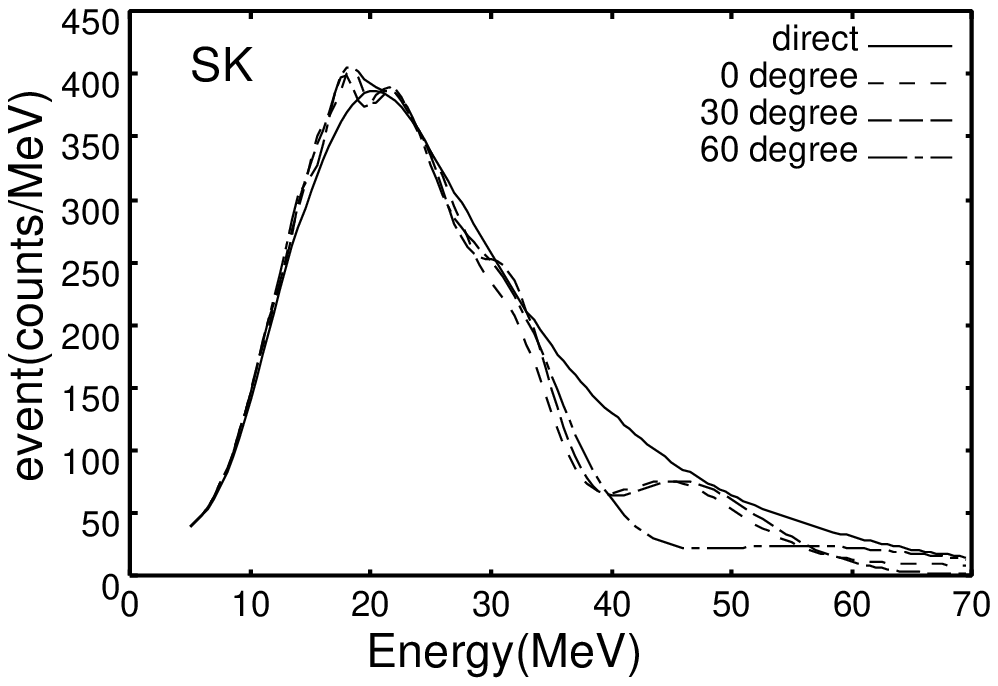}
\includegraphics[width=0.5\hsize]{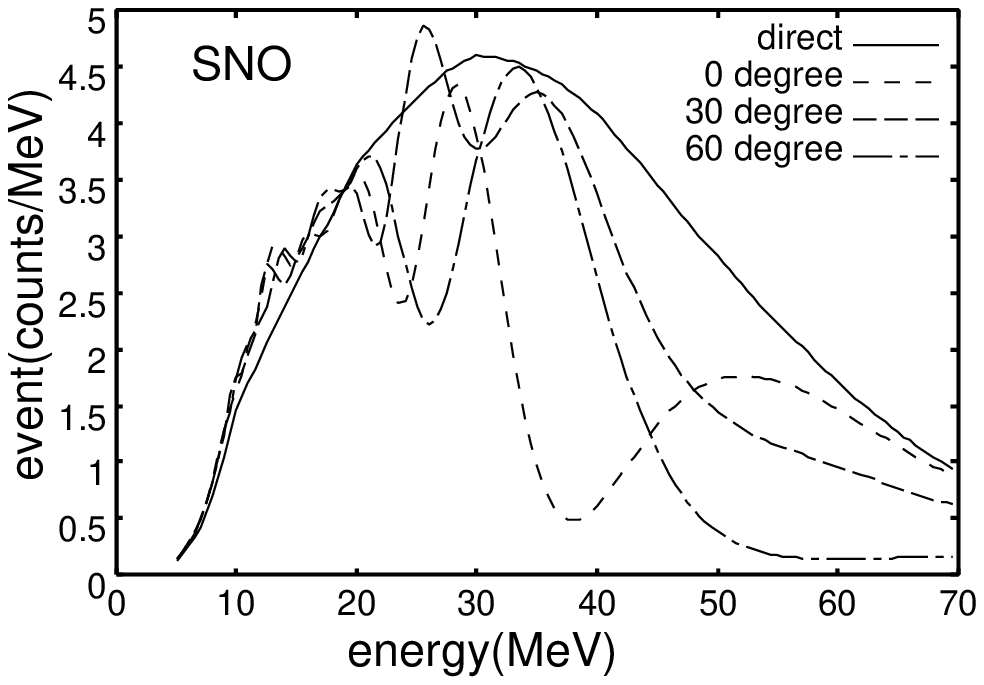}
\hss}
\caption{Earth effect on the SN spectra from fig.~\ref{fig:4} in
different detectors for different nadir angles. For SNO only CC events
are taken into account.  The oscillation parameters correspond to the
solar LMA case and a small $\theta_{13}$.  (Figures from
Ref.~\cite{Takahashi2001a} with permission.)}
\label{fig:6}
\end{figure}

Taking the numerical source model of Fig.~\ref{fig:4}, the
time-integrated spectra at Super-Kamiokande and SNO are shown in
Fig.~\ref{fig:6} for different nadir angles which determines the
Earth-crossing path. The oscillation parameters were chosen for the
LMA case with $\Delta m_{12}^2=2\times10^{-5}~{\rm eV}^2$, $\Delta
m_{13}^2=3.2\times10^{-3}~{\rm eV}^2$, $\sin^2\theta_{12}=0.87$, and
$\sin^2\theta_{23}=1.0$. The unknown third mixing angle was chosen
small as $\sin^2\theta_{13}=1.0\times10^{-6}$. Figure~\ref{fig:6}
illustrates that dramatic modifications of the spectra can be
expected for certain cases.

It is difficult to anticipate everything about future data. If a
galactic SN is observed, what we can learn about neutrino oscillations
depends on the detectors operating at that time and their geographical
location. It will also depend on the true source properties regarding
flavor-dependent spectra and fluxes, and what is already known about
the neutrino oscillation parameters as input information at that time.
Many authors have studied these questions
\cite{Takahashi2001a,Chiu1999,Dighe1999,Dutta2001,Fuller1998,%
Lunardini2001b,Minakata2000,Takahashi2001b,Takahashi2001c}.  It
appears that one may well distinguish between large and small values
of the elusive $\theta_{13}$ and between normal or inverted mass
hierarchies, or even accurately pin down $\Delta m_{12}^2$. An
important caveat is that these studies usually made unrealistic
assumptions about the source spectra and fluxes~\cite{Keil2002}. More
realistic cases remain to be investigated.  Still, a SN neutrino
observation would complement the upcoming efforts of precision
determination of neutrino oscillation parameters in long-baseline
experiments~\cite{Barger2000,Freund2001,Cervera2000}.


\section{Summary and Conclusions}

\label{sec:summary}

The compelling detection of flavor oscillations in the solar and
atmospheric neutrino data have triggered a new era in neutrino
physics. In the laboratory one will proceed with precision experiments
aimed at measuring the details of the mixing matrix. Future tritium
decay experiments may well be able to probe the overall neutrino mass
scale down to the 0.3~eV range, but if the absolute masses are
smaller, it will be very difficult to measure them, and the overall
mass scale may remain the most important unknown quantity in neutrino
physics for a long time to come.

Cosmological large-scale structure data at present provide the most
restrictive limit on neutrino masses of $\sum m_\nu<2.5~{\rm eV}$,
corresponding to $m_\nu<0.8~{\rm eV}$ in a degenerate mass scenario. A
rigorous relationship between the cosmic hot dark matter fraction
$\Omega_{\nu\bar\nu}$ and $m_\nu$ depends on the cosmic neutrino
density~$n_{\nu\bar\nu}$. If the solar LMA solution is correct,
big-bang nucleosynthesis constrains $n_{\nu\bar\nu}$ without further
assumptions about the neutrino chemical potentials. In the LMA case
neutrinos reach de-facto flavor equilibrium before the epoch of
weak-interaction freeze out.

While neutrinos provide only a small fraction of the cosmic mass
density, the mass and mixing schemes suggested by the oscillation
experiments are nicely consistent with leptogenesis scenarios for
creating the cosmic baryon asymmetry. Therefore, massive neutrinos may
be more closely related to the cosmic baryons than the dark matter.

Experimental neutrino physics is developing fast. It is foreseeable
that a broad range of neutrino detectors will operate for a long time,
perhaps decades, with a variety of goals from precision oscillation
physics to proton decay. Typically such detectors are also sensitive
to SN neutrinos so that one may expect the neutrino sky to be covered
perhaps for decades. Even though galactic SNe are rare, the neutrinos
from about a thousand SNe are on their way, and the expected time
scale of neutrino coverage makes a future high-statistics SN
observation a realistic possibility.

The existing neutrino mass limits are already so low that one can not
expect improving them with a galactic SN observation.  Turning this
around, we already know so much about neutrinos that we can be sure
that the observed SN neutrino light curve faithfully represents the
behavior of the source, without modifications by dispersion
effects. The same can not be said about the spectra which are
inevitably modified by flavor oscillations.  Depending on the details
of the source spectra, detection channel, and location on Earth,
flavor oscillations can have a large impact on the observations, and
may even allow one to disentangle some of the neutrino mixing
scenarios.

Arguably the most important benefit of a high-statistics measurement
of stellar collapse would be observing the dynamics of a cataclysmic
astrophysical event that could never be observed in any other
way. Whether or not numerical SN simulations will soon converge on a
theoretical standard model for the collapse and explosion mechanism,
the importance of its independent verification or falsification by a
detailed neutrino light curve can not be overstated.  As we learn more
about neutrinos, their future role is likely that of astrophysical
messengers, showing us phenomena that otherwise would remain forever
invisible.


\acknowledgments

This work was supported, in part, by the Deutsche Forschungsgemeinschaft
under grant No.\ SFB-375 and by the European Science Foundation (ESF)
under the Network Grant No.~86 Neutrino Astrophysics.


\end{document}